\documentclass[a4paper,fleqn,usenatbib]{mnras}

\usepackage[T1]{fontenc}
\usepackage{ae,aecompl}

\usepackage{graphicx}	% Including figure files
\usepackage{amsmath}	% Advanced maths commands
\usepackage{amssymb}	% Extra maths symbols

\title[Planet Hosts: Activity, Mass Loss and SPI]{SALT observations of the Chromospheric Activity of Transiting Planet Hosts: Mass Loss and Star Planet Interactions\footnotemark[1]\thanks{Based on
observations made with the Southern African Large Telescope (SALT), under programs 2013-2-UKSC-010, 2014-1-UKSC-OTH-001, 2014-2-SCI-049 (PI: C.A.Haswell) }}
\author[D. Staab, C.A. Haswell, Gareth D. Smith, L. Fossati, J.R. Barnes,  R. Busuttil, J.S. Jenkins]{D. Staab$^{1}$\thanks{E-mail:daniel.staab@open.ac.uk}, C.A. Haswell$^{1}$, Gareth D. Smith$^{1}$, L. Fossati$^{1,2}$, J.R. Barnes$^{1}$, 
\newauthor R. Busuttil$^{1}$ and J.S. Jenkins$^{3}$
\\
\\
$^{1}$Department of Physical Sciences, The Open University, Walton Hall, Milton Keynes, MK7 6AA, UK\\
$^{2}$Space Research Institute, Austrian Academy of Sciences, Schmiedlstrasse 6, Graz, A-8042, Austria\\
$^{3}$Departamento de Astronom\'ia, Universidad de Chile, Casilla 36-D, Las Condes, Santiago, Chile}

\date{Accepted 2016 December 2. Received 2016 December 1; in original form 2016 January 29}

\pubyear{2016}

\begin{document}
\label{firstpage}
\pagerange{\pageref{firstpage}--\pageref{lastpage}}
\maketitle

\begin{abstract}
We measured the chromospheric activity
of the four hot Jupiter hosts WASP-43, WASP-51/HAT-P-30, WASP-72 \& WASP-103 to
search for anomalous values caused by the close-in companions.
The Mount Wilson \hbox{Ca {\sc ii} H\,\&\,K} \textit{S}-index was calculated for each star using
observations taken with the
Robert Stobie Spectrograph at the Southern African Large Telescope.
The activity level of WASP-43 is anomalously high relative to its age and falls among the highest values
of all known main sequence stars.
We found marginal evidence that the activity of WASP-103 is also higher than expected from the system age.
We suggest that for WASP-43 and WASP-103 star-planet interactions (SPI) may enhance the \hbox{Ca {\sc ii} H\,\&\,K} core emission.
The activity levels of WASP-51/HAT-P-30 and WASP-72 are anomalously low,
with the latter falling below the basal envelope for both main sequence and evolved stars.
This can be attributed to
circumstellar absorption due to planetary mass loss, though absorption in the ISM may contribute.
A quarter of known short period planet hosts exhibit anomalously low activity levels, including systems with hot Jupiters
and low mass companions.
%Discrepancies between the activity level and isochronal age exist for all our targets.
Since SPI can elevate and absorption can suppress
the observed chromospheric activity of stars with close-in planets, their \hbox{Ca {\sc ii} H\,\&\,K} activity levels are an
unreliable age indicator.
Systems where the activity is depressed by absorption
from planetary mass loss are key targets for examining planet compositions through transmission spectroscopy.

\end{abstract}


\section{Introduction}

Stars with convective envelopes exhibit a wide range of magnetic activity levels, and consequently vary substantially in their
chromospheric emission.
The Mount Wilson program
(\citealt{Wilson1968}, \citealt{Duncan1991})
established the emission strength of the
\hbox{Ca {\sc ii} H\,\&\,K} line cores as the most widespread metric of chromospheric activity,
using the \textit{S}-index, a ratio between \hbox{Ca {\sc ii} H\,\&\,K}
emission flux and
continuum passbands.
A conversion from \textit{S}-values to $\log({\rm \textit{R}'_{HK}})$ \citep{Noyes1984} allows
consistent comparison of stars with differing spectral types
and has been measured for thousands of bright stars
(e.g. \citealt{Jenkins2011}; \citealt{Isaacson2010}; \citealt{Lovis2011}).
To first order, the stellar activity level depends on the stellar rotation rate and therefore age, with a particularly
rapid decrease during spin-down over the first $\sim$ 1 Gyr (e.g. \citealt{Mamajek2008}, hereafter MH08, and \citealt{Pace2013});
$\log({\rm \textit{R}'_{HK}})$ is thus used as a stellar age indicator.
%Large stellar samples tend to show a bimodal activity distribution with active and inactive subsamples, separated by the
%Vaughan-Preston gap (\citealt{Vaughan1980}; \citealt{Gray2006a}).

\citet{Knutson2010} published $\log({\rm \textit{R}'_{HK}})$ measurements for a sample of hot Jupiter (HJ) host stars, and concluded that
there was a correlation between stellar activity and the presence of a temperature inversion in the planetary atmospheres.
The latter was quantified with a model-independent metric based on the slope of the HJ secondary eclipse spectra.
\citet{Hartman2010} used the same sample to highlight a correlation between planetary surface gravity and $\log({\rm \textit{R}'_{HK}})$.
The effect of the activity
driven stellar high energy (XUV) flux on planetary atmospheres has been considered by e.g.
\citet{Lammer2003};
\citet{Erkaev2007};
\citet{Lopez2012};
\citet{Koskinen2013};
\citet{Jin2014};
\citet{Chadney2015}.
There has been significant interest in the chromospheric \hbox{Ca {\sc ii} H\,\&\,K} line flux in HJ host stars, with the hypotheses that:
\begin{enumerate}
\item
A HJ planet can stimulate stellar activity through magnetic and/or tidal star planet interactions (SPI; \citealt{Cuntz2000}).
\item
A HJ planet can suppress stellar activity through tidal interactions (\citealt{Miller2012}; \citealt{Pillitteri2014}).
\item
Mass loss from a HJ planet can form a diffuse circumstellar gas cloud which absorbs in the cores of strong resonance lines (e.g. \hbox{Ca {\sc ii} H\,\&\,K} and \hbox{Mg {\sc ii} h\,\&\,k}) suppressing the measured stellar activity below its true value (\citealt{Haswell2012}; \citealt{Fossati2013}).
\end{enumerate}

It is important to identify individual systems where mass loss
appears to be masking the intrinsic activity, and cases where the stellar activity appears boosted by SPI.
\hbox{Ca {\sc ii} H\,\&\,K} emission enhancements from SPI should be a good predictor of a system's
radio brightness (e.g. \citealt{See2015}). With dramatic enhancements in radio astronomy capabilities
(ALMA, \citealt{Partnership2015}; LOFAR, \citealt{VanHaarlem2013}; HERA, \citealt{Hewitt2011}; SKA, \citealt{Carilli2004}), detections
of exoplanetary radio emissions are imminent \citep{Vidotto2015}. Radio observations are expected to yield planet rotation periods and magnetic
moments, with important implications for exoplanetary magnetospheric physics and the transfer of energy and angular momentum between
the host star and planet.

Systems where the \hbox{Ca {\sc ii} H\,\&\,K} emission is absorbed by gas lost from the planet offer the potential to determine
 the planet's chemical composition through transmission spectroscopy. In WASP-12, while the diffuse gas is present at all observed phases, it produces
the most near-UV absorption close to transit \citep{Haswell2012}. Differencing the observed spectrum near transit and away
from transit can reveal the additional absorption from the densest regions of the gas shroud. Because the near-UV
is such an informative wavelength region, with strong resonance lines of many abundant elements and ions, it is vital to make
Hubble Space Telescope observations of the shrouded systems. There is no alternative means to obtain comparable information on
planetary composition.

There are few southern hemisphere spectrographs calibrated to produce $\log({\rm \textit{R}'_{HK}})$, and large telescopes or significant exposure times are required to achieve sufficient signal to noise in the \hbox{Ca {\sc ii} H\,\&\,K} cores of typical HJ host stars (V= 11 or fainter).
We report observations using the Robert Stobie Spectrograph (RSS) at the
Southern African Large Telescope (SALT), calibrating it to measure $\log({\rm \textit{R}'_{HK}})$, and results for four HJ host stars. Section~\ref{Section:ObsRedux} describes observations and data reduction; Section~\ref{Section:analysis} describes the calibration; Section~\ref{Section:Discussion} discusses our planet host measurements in the context of large stellar samples and the three hypotheses listed above; Section~\ref{Section:Conclusion} gives our conclusions and implications for future work.
%, discusses our results and previously measured $\log({\rm \textit{R}'_{HK}})$ values for planet hosts examining .

\section[]{Observations and Reduction} \label{Section:ObsRedux}
Our observations were taken with the Robert Stobie Spectrograph (RSS; \citealt{Kobulnicky2003}), a multimode instrument at the Southern African Large Telescope
(SALT; \citealt{Buckley2006}).
%It employs a set of slit masks and volume phase holographic transmission gratings and an articulating camera/detector system,
%allowing a choice of wavelength coverage and spectral resolution.
\citet{Jenkins2011} assessed the effect of spectral resolution on the precision of $\log({\rm \textit{R}'_{HK}})$ measurements, finding a resolving power of R $>$ 2500 is needed.
We selected RSS settings to achieve the highest possible resolution in the region of the \hbox{Ca {\sc ii} H\,\&\,K} lines:
a 0.6\arcsec slit, and the PG3000 grating, with
%To place the wavelength range of interest away from detector gaps, we
%chose
a camera station angle of 79.75\degr \,and grating angle of 39.875\degr.
This yielded coverage from 3882\,\AA{} to 4614\,\AA{}, and a resolution R $\sim$ 7300 at 4000 \AA{}.
Prebinning of 2 in the spectral direction gave 0.23 \AA{} per binned pixel.
Future $\log({\rm \textit{R}'_{HK}})$ measurements with the RSS using our calibration
should adopt the same setup to ensure consistency.

%\subsection{Target selection}

To permit calibration on the Mount Wilson system, we selected calibration stars
from \citet{Baliunas1995}, hereafter B95. B95 reported both the variability and the mean of \textit{S}-values, $<S_{\rm{MW}}>$,
using decades of data from the Mount Wilson survey.
We included stars ranging from very inactive,
$<S_{\rm{MW}}>$ = 0.14, to very active, $<S_{\rm{MW}}>$ = 0.5, (Table~\ref{Table:Calibrators}),
 chosen for
low variability and $ 0.4 < B-V < 1.0$.
%, and to be observable with SALT.
HD\,10700 and HD\,22049 showed stable activity for a decade after B95's observations \citep{Hall2007}.
%\citet{Hall2007} showed that their activity was stable at the B95 level for another decade.
%The calibrator stars are bright, so
We took 4 - 6 consecutive, short (15s - 70s), high SNR exposures, to avoid  saturation of our calibration star spectra.

\begin{figure}
    \includegraphics[width=85mm]{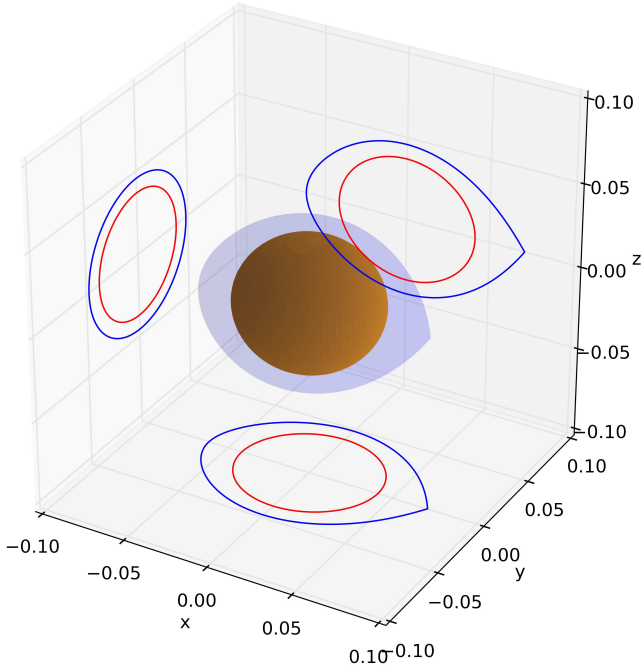}
    \caption{The Roche lobe (blue) of WASP-103b (orange), calculated using stellar and planetary parameters from TEPCAT \citep{Southworth2011}. We corrected
the empirical radius for the projection effect of a non-spherical planet viewed at orientation $i \ne 90^{\circ}$, to
give a self-consistent 3-D Roche geometry. Axes are in units of orbital separation, with the x and y axis in the orbital plane and z perpendicular to it.
For the projections, red outlines represent the planet and blue outlines the Roche lobe.}
    \label{Fig:Roche}
\end{figure}

Our four science targets, WASP-43, WASP-51/HAT-P-30, WASP-72, and WASP-103 (\citealt{Hellier2011}; \citealt{Johnson2011}; \citealt{Gillon2013};
 \citealt{Gillon2014}; Table~\ref{Table:BasicData}),  are host stars of transiting HJ planets. They were all discovered after 2010 and are notable for
%selected on the basis of discovery after the
%publication of \citet{Knutson2010},
proximity to their host stars.
%, and visibility to SALT.
%Our targets are  Basic stellar and planetary parameters are given in .
%In Fig.~\ref{Fig:Roche} we show the Roche geometry for these four planets; we will discuss
%this figure in more detail in Section~\ref{sec:Roche}.
WASP-103\,b
%suffers appreciable tidal deformation due to proximity to the host star,
%and
almost fills its Roche lobe (Figure~\ref{Fig:Roche});
%and
%. In this respect, WASP-103b is similar to WASP-12b (c.f. Figure 21 of \citealt{Haswell2012}) and
we
%chose to
made multiple
exposures of this target to  search for possible variations in circumstellar absorption of  \hbox{Ca {\sc ii} H\,\&\,K}.
%line cores with orbital
%phase.
%To help with the already highly constrained SALT scheduling, we did not specify particular observing phases, but allowed our visits to WASP-103 to
%fall as determined by the unconstrained queue-scheduling.

\begin{table*}
 \centering
 \begin{minipage}{140mm}
  \caption{Stellar and planetary parameters relevant for this work, taken from TEPCAT \citep{Southworth2011}, including stellar effective temperature ($T_{\rm{eff}}$),
semi-major axis ($a$), orbital period and planetary equilibrium temperature ($T_{\rm{eq}}$).}
  \begin{tabular}{@{}llllllll@{}}
  \hline
   System &  $T_{\rm{eff}}$ (K) & $a$ (AU) &  Period (d) & $M_{\rm{p}}\ (M_{\rm{J}})$   & $R_{\rm{p}}\ (R_{\rm{J}})$   & $g_{\rm{p}}$ ($ms^{-2}$)   & $T_{\rm{eq}}$ (K)    \\
  \hline
  HAT-P-30/WASP-51 & 6338 $\pm$ 42 & 0.042 & 2.81 & 0.71 $\pm$ 0.03 & 1.34 $\pm$ 0.07 & 9.8 $\pm$ 0.9 &  1630 $\pm$ 42 \\
  WASP-43  & 4520 $\pm$ 120 & 0.015 & 0.81 & 2.03 $\pm$ 0.05 & 1.04 $\pm$ 0.02 & 47.0 $\pm$ 1.4 &  1440 $\pm$ 40 \\
  WASP-72  & 6250 $\pm$ 100 & 0.037 & 2.23 & 1.46 $\pm$ 0.06 & 1.27 $\pm$ 0.20   & 22.9 $\pm$ 7.3 &  2210 $\pm$ 120 \\
  WASP-103 & 6110 $\pm$ 160 & 0.020 & 0.93 & 1.47 $\pm$ 0.11  & 1.55 $\pm$ 0.05 & 15.1 $\pm$ 0.9 &  2495 $\pm$ 66 \\

\hline
\end{tabular}
\label{Table:BasicData}
\end{minipage}
\end{table*}

Tables~\ref{Table:Calibrators} and~\ref{Table:HostActivities} summarise the observations. The calibrators HD\,26913 and HD\,26923 are separated by $\sim$ 40 arcsec, and were therefore observed simultaneously by appropriately orientating
the RSS slit.
The SALT pipeline \citep{Crawford2010} performs corrections for CCD bias, gain and crosstalk between the CCD amplifiers.
We performed flat fielding, background subtraction and wavelength calibration in IRAF, using arc-lamp exposures taken immediately after each target.
Spectra were optimally extracted with particular attention to background subtraction and shifted into the stellar rest frame by cross-correlation with the
National Solar Observatory solar spectrum \citep{Kurucz1984},
degraded to the RSS spectral resolution.
All spectra have SNR $>$ 15 in the \hbox{Ca {\sc ii} H\,\&\,K} line cores. We rebinned all spectra to a common wavelength scale with 0.05 \AA{} bin width,
while conserving flux.

\section{Analysis} \label{Section:analysis}
\subsection{Extraction of instrumental \textit{S}-values}

\begin{figure}
\includegraphics[width=85mm]{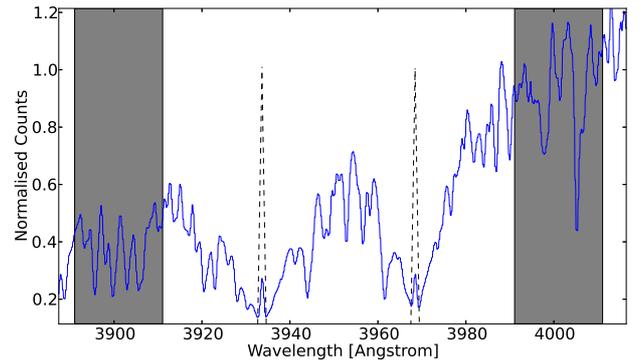}
\caption{An RSS spectrum of our most active calibrator star, HD\,22049, with continuum windows (greyed) and triangular core bandpasses
highlighted. Strong \hbox{Ca {\sc ii} H\,\&\,K} emission cores are evident. Counts are normalised to the mean of the red continuum bandpass.}
\label{Fig:BandPassILLU}
\end{figure}

We
%replicated the methodology of the Mount Wilson spectrophotometer
performed synthetic photometry using the \hbox{Ca {\sc ii} H\,\&\,K} core bandpasses, $H$ and $K$, and continuum windows, $R$ and $V$,
shown in Fig.~\ref{Fig:BandPassILLU}. $V$ and $R$ are the mean flux values in 20 \AA{} wide continuum windows centered on 3901.07 \AA{} and 4001.07 \AA{} while
$H$ and $K$ are the core bandpasses, centered
on 3933.664 \AA{} and 3968.47 \AA{}, triangularly weighted with a FWHM of 1.09 \AA{}. %(Fig.~\ref{Fig:BandPassILLU}).
We followed \citet{Lovis2011} in using the (weighted) mean bandpass fluxes.
This reduces edge effects from finite pixel size at the band boundaries, and makes a historical scaling factor obsolete.
The instrumental \textit{S}-index, $S_{\rm{RSS}}$, is given by
\begin{equation}
      S_{\rm{RSS}} = \frac{H \ + \ K}{R \ + \ V} \ ,
\end{equation}
Note flux-calibration is unnecessary for \textit{S}-index
measurements: the bandpass window placement makes these insensitive to the local spectral slope (e.g. \citealt{Gray2003}).

%
% on the reduced spectra, as done by the majority of activity surveys in the literature.
%The \hbox{Ca {\sc ii} H\,\&\,K} core bandpasses and continuum windows either side
%are shown in Fig.~\ref{Fig:BandPassILLU}.

%\citet{Pepe2011} advocate a robust and simple approach, calculating the mean flux
% each bandpass rather than using integrated fluxes. This reduces any edge effects from finite pixel size at the band boundaries.
%To further mitigate such effects we rebinned all spectra to a common wavelength scale with 0.05 \AA{} bin width,
%while conserving flux. The use of mean fluxes also makes a historical scaling factor obsolete, which appears in other conversions
%to the Mount Wilson system \citep{Pepe2011}.

\subsection{Calibration to Mount Wilson system}

\begin{figure}
\includegraphics[width=85mm]{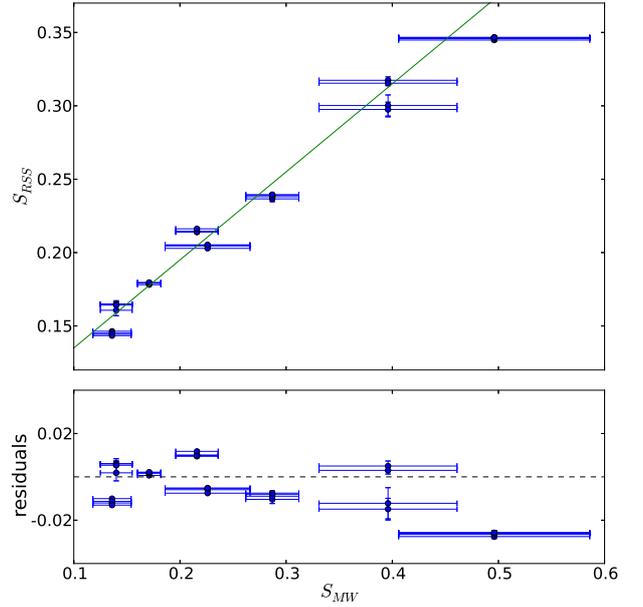}
\caption{Calibration from instrumental to Mount Wilson \textit{S}-values. Individual datapoints for each calibrator star overlap closely, and
in most cases have $S_{\rm{RSS}}$ uncertainties smaller than the symbol sizes.}
\label{Fig:CalibrationRelation}
\end{figure}

Table~\ref{Table:Calibrators} reports the calibration measurements with propagated photon noise uncertainties on $S_{\rm{RSS}}$.
For the B95 values,
formal measurement uncertainty on $<S_{\rm{MW}}>$  is negligible compared to astrophysical stellar activity variation.
The current activity level of our calibrators is uncertain so we report the variability range measured over the multi-decade baseline of B95 in
Table~\ref{Table:Calibrators} and Figure~\ref{Fig:CalibrationRelation}.
The activity variability observed by B95
increases with the mean stellar activity level.

\begin{table}
% \centering
% \begin{minipage}{80mm}
  \caption{Calibrator star measurements and associated photon noise uncertainties. Mean Mount Wilson \textit{S}-values and their variability range were taken from B95. Note
that HD\,26913 and HD\,26923 were observed simultaneously.}
  \begin{tabular}{@{}lllc@{}}
  \hline
   Name     &    $< S_{\rm{MW}} >$ &  $S_{\rm{RSS}}$   & HJD-2450000   \\
  \hline
HD\,10700 &   0.171 $\pm$ 0.011 &  0.1781 $\pm$ 0.0001 &     6658.339\\
    & &    0.1791 $\pm$   0.0001 &     6658.340 \\
   & &    0.1794  $\pm$  0.0001   &   6658.341\\
   & &    0.1796  $\pm$  0.0001    &  6658.342\\
  HD\,182101 &  0.216 $\pm$ 0.020 &  0.2139 $\pm$ 0.0004   &   6792.586\\
   & &   0.2145  $\pm$  0.0004     & 6792.587\\
   & &   0.2143  $\pm$  0.0004     & 6792.587\\
   & &   0.2162  $\pm$  0.0004     & 6792.588\\
  HD\,9562 &  0.136 $\pm$ 0.018 & 0.1444 $\pm$ 0.0002  &    6652.321\\
   & &  0.1464 $\pm$ 0.0002     & 6652.322\\
   & &  0.1451 $\pm$ 0.0002     & 6652.323\\
   & &  0.1434 $\pm$ 0.0002     & 6652.324\\
  HD\,22049  &  0.496 $\pm$ 0.090 &  0.3465 $\pm$ 0.0009   &   6984.514\\
   & &  0.3464 $\pm$ 0.0010     & 6984.514\\
   & &  0.3465 $\pm$ 0.0011     & 6984.515\\
   & &  0.3448 $\pm$ 0.0009     & 6984.515\\
   & &  0.3458 $\pm$ 0.0006     & 6984.515\\
   & &  0.3466 $\pm$ 0.0007     & 6984.516\\
  HD\,26913  &  0.396 $\pm$ 0.065 & 0.3002 $\pm$ 0.0072   &   6985.407\\
   & &  0.2975 $\pm$ 0.0050     & 6985.407\\
   & &  0.3174 $\pm$ 0.0023     & 6985.408\\
   & &  0.3154 $\pm$ 0.0017     & 6985.409\\
  HD\,26923  &  0.287 $\pm$ 0.025 & 0.2390 $\pm$ 0.0015   &   6985.407\\
   & &  0.2366 $\pm$ 0.0019      &6985.407\\
   & &  0.2381 $\pm$ 0.0010      &6985.408\\
   & &  0.2395 $\pm$ 0.0006     & 6985.409\\
  HIP\,110785 & 0.140 $\pm$ 0.015 & 0.1649 $\pm$ 0.0013   &   6803.656\\
   & &  0.1649 $\pm$ 0.0011      &6803.656\\
   & &  0.1642 $\pm$ 0.0030      &6803.657\\
   & &  0.1607 $\pm$ 0.0037      &6803.658\\
  HIP\,12114 & 0.226 $\pm$ 0.040   &0.2050 $\pm$ 0.0005   &   6627.328\\
   & &  0.2045 $\pm$  0.0005    &  6627.328\\
   & &  0.2029 $\pm$  0.0005     & 6627.329\\
   & &  0.2053 $\pm$  0.0005      &6627.330 \\

\hline
\end{tabular}
\label{Table:Calibrators}
%\end{minipage}
\end{table}

Following \citet{Jenkins2011}, we perform a simple linear calibration to the Mount Wilson system, using ordinary least squares fitting (Figure~\ref{Fig:CalibrationRelation}) obtaining:
\begin{equation}
S_{\rm{RSS}} = (0.60 \pm 0.02) \ S_{\rm{MW}} + (0.075 \pm 0.005).
\end{equation}
The RMS of the fit, 0.013, is at the level found in calibrations of other instruments (see \citealt{Jenkins2006}; \citealt{Jenkins2008}; \citealt{Arriagada2011}).
Scatter at the $\sim$~5~\% level is an unavoidable consequence of the long-term activity variation of calibration stars.
% combined with the single-epoch measurements
% of their $S_{\rm{RSS}}$ values.
Note that the high activity calibrators carry larger uncertainties and consequently less weight in the
fit.

\begin{figure}
\includegraphics[width=85mm]{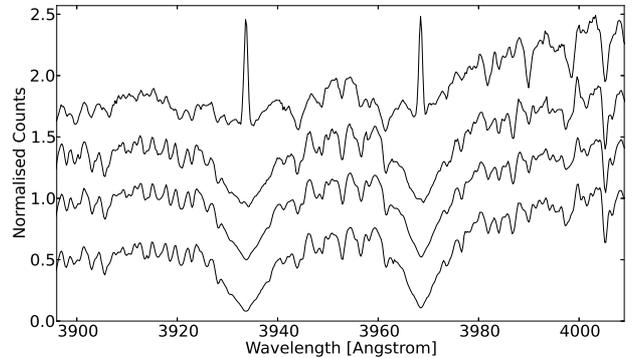}
\caption{Spectra plotted in order of decreasing apparent activity from top to bottom: WASP-43, WASP-103, WASP-51, WASP-72.
An arbitrary offset was added to the spectra for visibility. Counts are normalised to the mean of the red continuum bandpass.}
\label{Fig:TargetSpectra}
\end{figure}

Table~\ref{Table:HostActivities} lists $S_{\rm{MW}}$ for the planet hosts,
calculated from Equations 1 and 2,
and Figure~\ref{Fig:TargetSpectra} shows the \hbox{Ca {\sc ii} H\,\&\,K} lines of our planet hosts.
For the final conversion to $\log({\rm \textit{R}'_{HK}})$ following \citet{Noyes1984}, the stellar B-V is required.
No B-V measurements
% has not been directly measured
for WASP-103 have been published;
% in the literature, and both
Simbad\footnote{http://simbad.u-strasbg.fr/simbad/} and
the Exoplanet Orbit Database\footnote{http://exoplanets.org/} \citep{Han2014} report
B-V colors for WASP-72 and WASP-51 that are completely inconsistent with detailed spectral analyses in the discovery papers
%their spectral types and effective
%temperatures
(\citealt{Gillon2013} and \citealt{Johnson2011}).
%Note that the latter parameters are from detailed analyses in the discovery papers of the respective planets.
%For consistency,
We calculated B-V for our planet hosts (Table \ref{Table:HostActivities}) from the stellar parameters reported in
TEPCAT, using Equation 3 in
\citet{Sekiguchi2000}  propagating all uncertainties.
% on the relevant stellar data.

\begin{table*}
 \centering
 \begin{minipage}{180mm}
  \caption{Planet host measurements and activity values derived from them. Note instrumental \textit{S}-values report photon noise uncertainties,
while $S_{\rm{MW}}$ includes the calibration uncertainty. B-V values and uncertainties are derived as
described in Section~\ref{Section:analysis}. Orbital phases were calculated from ephemerides in TEPCAT, with phase 0 representing
mid-transit. Stellar ages in the literature are compared with activity ages derived from the MH08 relation, using our $\log({\rm \textit{R}'_{HK}})$ values.}
  \begin{tabular}{@{}lllllllcll@{}}
  \hline
   Name  & HJD$_{-2450000}$ & phase & $t_{\rm{exp}}$(s) &$ S_{\rm{RSS}} $ &  $S_{\rm{MW}}$ & B-V & $\log({\rm \textit{R}'_{HK}})$ & activity age\footnote{Uncertainty propagated
from uncertainty in  $\log({\rm \textit{R}'_{HK}})$. Note MH08 reports activity-age scatter at the 60\% and 30 \% levels for ages below and above 130 Myr
respectively.}  & literature age\footnote{(1)\citealt{Hellier2011},(2)\citealt{Johnson2011}, (3)\citealt{Bonfanti2015}, (4)\citealt{Gillon2013}, (5)\citealt{Southworth2014}, (6)\citealt{Gillon2014}}  \\
  \hline
WASP-43 & 6689.420 &0.661& 2032   &1.209 $\pm$ 0.008 &  1.889 $\pm$ 0.065  & 1.10 $\pm$ 0.06 & -4.17 $\pm$ 0.10 & 45 $\pm$ 40 Myr & 300-600 Myr(1)   \\
WASP-51 & 6698.410 &0.883& 1108  &0.168 $\pm$ 0.001 &  0.155 $\pm$ 0.010 & 0.49 $\pm$ 0.01 & -4.98 $\pm$ 0.07 & 6.2 $\pm$ 1.3 Gyr & 0.5-1.8 Gyr(2)    \\
                                                                                                                         &&&&&&&&& 0.6-1.0 Gyr(3)    \\
WASP-72 & 6606.307 &0.332& 1108  &0.149 $\pm$ 0.001 &  0.124 $\pm$ 0.009 & 0.48 $\pm$ 0.03 & -5.30 $\pm$ 0.15 &  11 $\pm$ 0.8 Gyr & 2.6-3.8 Gyr(4)      \\
WASP-103& 6779.562 &0.701& 317   &0.224 $\pm$ 0.003 & 0.248 $\pm$ 0.013 & 0.54 $\pm$ 0.05 & -4.59 $\pm$ 0.04 & 950 $\pm$ 230 Myr\footnote{from mean of our measurements} & 1.8-6.2 Gyr(5)     \\
 & 6779.566&0.706& 317   & 0.221  $\pm$   0.003  &    0.244 $\pm$ 0.013 & & -4.60 $\pm$ 0.04 & &  3-5 Gyr(6)  \\
 & 6779.573&0.713& 317  & 0.221  $\pm$   0.003  &    0.243 $\pm$ 0.013 & &-4.60 $\pm$ 0.04 & &  \\
 & 6780.489&0.703& 317  & 0.227  $\pm$   0.003   &   0.253 $\pm$ 0.013 & &-4.57 $\pm$ 0.04 & & \\
 & 6780.493&0.707& 317  & 0.227  $\pm$   0.003   &   0.254 $\pm$ 0.013 & &-4.57 $\pm$ 0.04  & & \\
 & 6780.497&0.712& 317  & 0.233  $\pm$   0.003   &   0.263 $\pm$ 0.013 & &-4.55 $\pm$ 0.03 & &  \\
 & 6780.501&0.716& 317  & 0.230  $\pm$   0.003   &   0.258 $\pm$ 0.013 & &-4.56 $\pm$ 0.04  & &  \\
 & 6780.535&0.753& 390  & 0.220  $\pm$   0.009   &   0.241 $\pm$ 0.019 & &-4.61 $\pm$ 0.06 & &  \\
 & 6790.473&0.490& 390 & 0.228  $\pm$   0.003   &   0.256 $\pm$ 0.013 & &-4.57 $\pm$ 0.04 & &  \\
 & 6790.478&0.496& 390  & 0.229  $\pm$   0.003   &   0.257 $\pm$ 0.013 & &-4.56 $\pm$ 0.04 & &  \\
 & 6790.482&0.500& 390  & 0.235  $\pm$   0.003   &   0.267 $\pm$ 0.013 & &-4.54 $\pm$ 0.03 & &  \\
 & 6790.487&0.505& 390  & 0.237  $\pm$   0.003   &   0.269 $\pm$ 0.013 & &-4.53 $\pm$ 0.03 & &  \\
 & 6790.502&0.521& 390  & 0.227  $\pm$   0.003   &   0.254 $\pm$ 0.013 & &-4.57 $\pm$ 0.04 & &   \\
 & 6790.507&0.527& 390  & 0.234  $\pm$   0.003   &   0.264 $\pm$ 0.013 & &-4.54 $\pm$ 0.03 & &  \\
 & 6790.511&0.531& 390 & 0.225  $\pm$   0.003   &   0.251 $\pm$ 0.013 & &-4.58 $\pm$ 0.04 & & \\
 & 6790.516&0.537& 390  & 0.231  $\pm$   0.003   &   0.260 $\pm$ 0.013 & &-4.55 $\pm$ 0.03 & &  \\
\hline
\end{tabular}
\label{Table:HostActivities}
\end{minipage}

\end{table*}

\subsection{Uncertainty Budget}

%As shown by Table~\ref{Table:Calibrators} and Figure~\ref{Fig:CalibrationRelation},
The $S_{\rm{RSS}}$ values measured for separate exposures of the same calibrator star
agree to within 1\%, with the single exception of 3\% for HD\,26913.
%This demonstrates that our reduction and
% extraction steps are robust, and indicates excellent instrumental stability, at least on short timescales.
%For the brightest calibrator stars we reached
With SNR $\sim$ 1500 at 4000 \AA{}, there is
%leading to
a very small photon noise contribution
to the calibration star $S_{\rm{RSS}}$ values.
When we divide the RMS of the $S_{\rm{RSS}}$ values for each calibrator by the corresponding mean photon
noise, a sharp rise above unity occurs below
a photon noise level of $\sim$ 0.2 \%
due to a combination of
astrophysical variability in the chromospheric emission and systematic uncertainties.
%
% . This noise floor could be ,
%as the processes powering  are known to be variable on short timescales \citep{Narain1996}.
%Systematic uncertainties from instrumental stability and reduction steps may also be a plausible
%explanation. To put our precision in context, note that \citet{Pepe2011} reached a 7-year timescale \textit{S}-index precision of 0.35 \% for $\tau$\,Ceti
%HD\,10700
%with the ultra-stable
% HARPS spectrograph.

Our $S_{\rm{MW}}$ (Table~\ref{Table:HostActivities}) have uncertainties of 3 - 8\%, largely dominated by the uncertainty from the calibration relation.
Clearly the noise floor contribution of $\sim$ 0.2 \% is negligible.
The $\log({\rm \textit{R}'_{HK}})$ values listed in Table~\ref{Table:HostActivities} incorporate the propagated errors on  B-V, a significant contribution
for WASP-43 which has a poorly constrained effective temperature.
Obviously our $\log({\rm \textit{R}'_{HK}})$ values are snapshots during activity cycles and
short-term variability. \citet{DaSilva2013} used a sample of 271 mostly inactive stars, obtained over a $\sim$ 10 year timescale,
to show that astrophysical variation in $\log({\rm \textit{R}'_{HK}})$ results in an RMS of typically 0.015 dex, and at most 0.08 dex.
Our uncertainties render the typical astrophysical variability negligible.

%,
%rather than photon noise. Clearly the noise floor contribution of $\sim$ 0.2 \% is negligible.
%For the conversion of $S_{\rm{MW}}$ to $\log({\rm \textit{R}'_{HK}})$ values listed in Table~\ref{Table:HostActivities}, we propagated the error on
%our B-V values. This has a negligible effect for all
%targets except WASP-43, which has a relatively poorly constrained effective temperature.
%It should be noted that In comparison to our uncertainties in Table~\ref{Table:HostActivities}, the typical astrophysical variability is negligible.
%If future observations do show significant $\log({\rm \textit{R}'_{HK}})$ variability, this should be
%accounted for in the uncertainty of the mean activity level.

\section{Discussion}\label{Section:Discussion}
Interest in $\log({\rm \textit{R}'_{HK}})$ for individual planet hosts is motivated by evidence it might provide regarding magnetic and tidal star-planet interactions and
by the implications of the anomalously low values of $\log({\rm \textit{R}'_{HK}})$ of some HJ hosts stars.

\subsection{Star-Planet Interactions}
Stellar activity \textit{enhancements} have been detected for individual HJ systems
(e.g. \citealt{Shkolnik2008}; \citealt{Pillitteri2011}) and around periastron for a very eccentric system \citep{Maggio2015}. Null results for both chromospheric and coronal emission enhancements can be found in e.g.
\citet{Poppenhaeger2010}, \citet{Lenz2011} and \citet{Figueira2016}.
%detail cases of transient, ``on/off'' SPI, leading to  \hbox{Ca {\sc ii} H\,\&\,K} emission
%increases in phase with the planetary orbit at some epochs.
Activity \textit{suppression} by a planet has been suggested by \citet{Miller2012} and \citet{Pillitteri2014} to explain the anomalously low X-ray and
\hbox{Ca {\sc ii} H\,\&\,K} line core emission of WASP-18.
The parameters of this system, $M_{P} = 10.4 \ M_{\rm{J}} $, $P = 0.9$ d, suggest a particularly strong tidal interaction.
The magnitude of this effect should be greater than for any other HJ system, and may significantly affect the shallow stellar convective zone of the F6V host,
depressing its magnetic dynamo \citep{Pillitteri2014}.

Several studies have investigated whether mean stellar activity levels are systematically influenced by the presence and properties of planetary companions
across larger stellar populations.
Results have been very mixed for both X-ray measurements (e.g. \citealt{Scharf2010} in comparison with
\citealt{Poppenhaeger2011}) and  \hbox{Ca {\sc ii} H\,\&\,K} data (e.g. \citealt{Martins2011} in comparison with \citealt{Krejcova2012}).
It seems that at best optical and X-ray diagnostics reveal magnetic SPI effects in only a small
subset of observations of short period planets.
Tidal spin-up of host stars by orbital decay of HJs could play a more important, but nonetheless, limited role \citep{Miller2015}.
The latter process would `rejuvenate' the parent star, delaying or reversing the decline of rotation and activity with age.
For a comprehensive overview of SPI studies and the most recent statistical work to date, see \citet{Miller2015}.
SPI detection in large samples of systems remains controversial and beset by selection biases. However, strong evidence for activity enhancements
in individual cases has been found in particular for planet-hosting wide binary systems \citep{Poppenhaeger2014a}.
Theoretical work on magnetic SPI effects ranges from simple analytical approaches (e.g. \citealt{Lanza2008}) to more sophisticated, three-dimensional MHD models
(e.g. \citealt{Strugarek2015}). Studies such as \citet{Saur2013} and \citet{Lanza2012} cannot account for the energy release observed for time-variable SPI signatures.
Treatments which consider the dynamical behaviour of the magnetic field topology allow the phenomenon of magnetic reconnection to be included.
\citet{Lanza2012} finds that the power dissipated by reconnection between stellar and planetary fields at the planet's magnetospheric boundary is
insufficient to explain observations attributed to SPI. In contrast, relaxation of stressed magnetic loops between stellar and planetary fields can provide
sufficient power \citep{Lanza2013}.
\citet{Cohen2011} perform time-dependent MHD modelling, concluding that there is sufficient energy release from reconnection
to explain observed SPI effects. These studies also provide explanations for the intermittent nature of magnetic SPI signatures seen in e.g. \citet{Shkolnik2008}.
However, both observational and theoretical work on SPI remain active areas of research without conclusive outcomes to date.

\subsection{Circumstellar Absorption of Chromospheric Emission}

Large-scale hydrodynamic escape of the upper atmosphere
of irradiated short-period planets is predicted by models (\citealt{Lammer2003}; \citealt{Bisikalo2013}; \citealt{Matsakos2015}) and has been directly detected through transmission
spectroscopy. Notable cases are the HJs HD\,209458b \citep{Vidal-Madjar2003} and HD\,189733b (\citealt{LecavelierdesEtangs2010};
\citealt{Bourrier2013})
and the spectacular case of the warm Neptune GJ 436b (\citealt{Kulow2014}; \citealt{Ehrenreich2015}). %In addition to prodigious hydrogen,
Heavy species are
entrained in these prodigious outflows of hydrogen (\citealt{Linsky2010}; \citealt{Fossati2010}; \citealt{Haswell2012}; \citealt{Ben-Jaffel2013}).
\citet{Haswell2012} suggested that these outflows can feed diffuse circumstellar gas shrouds which absorb the stellar flux in the cores of strong lines of abundant species, depressing the chromospheric emission which arises in precisely these spectral lines. In particular this naturally explains the anomalous zero flux
\hbox{Mg {\sc ii} h\,\&\,k} line cores observed in the extreme HJ host WASP-12. Corroborating this, \citet{Fossati2013} showed that WASP-12's
\hbox{Ca {\sc ii} H\,\&\,K} lines have an extremely low value of $\log({\rm R'_{HK}}) = -5.5$.
If intrinsic to the star, this would be unique: WASP-12 is the lowest point shown in each panel of Figure~\ref{Fig:Test1}. Occam's razor suggests this extreme observed property of the star WASP-12 must be related to the presence of its extreme HJ planet, WASP-12b.

This \textit{apparent activity suppression} arises from planetary mass loss which is sensitive to the planet's surface gravity.
There is a highly significant correlation between planetary
surface gravity ($g_{\rm{p}}$) and stellar $\log({\rm \textit{R}'_{HK}})$ for close-in planets \citep{Hartman2010}. \citet{Figueira2014} confirmed this,
%with a 3 times larger sample, and
showing it is not due to selection biases.
\citet{Lanza2014} constructed a physical model reproducing the observed correlation, which was refined by
\citet{Fossati2015a}.
Large-scale planetary mass loss akin to that inferred from apparent activity suppression may underly the sub-Jovian desert
(\citealt{Kurokawa2014}; \citealt{Lundkvist2016}). Other explanations of this dearth in the known exoplanet population have been suggested however 
(\citealt{Matsakos2016} and references therein).

\subsection{Activity values in context} \label{Section:ActivityContext}

%\subsubsection{Comparison with Field Stars and other Transiting Planet Hosts}

%Our results are in broad agreement with the observed correlations between stellar activity and planetary surface gravity.

%Figure~\ref{Fig:Surfacegravity} shows our results in the context of the observed correlations between stellar activity and planetary surface gravity
%for a subset of transiting planet-hosts, as described in \citet{Fossati2015a}.
%Note that membership of the active and inactive subsets is assigned using an a posteriori probability criterion,
%and not with a specific activity cut-off level.
%The archival activity distribution of \citet{Fossati2015a} is broadly in agreement with our values; we will discuss individual targets in Section~\ref{Section:IndividualSystems}.

%\begin{figure}
%\includegraphics[width=85mm]{surfaceGravity-logRHKplot-viaTEPCAT-withAGEpoint.ps}
%\caption{Updated plot of $\log({\rm \textit{R}'_{HK}})$ against the inverse of planetary surface gravity ($g_{\rm{p}}$) from \citet{Fossati2015a}, with their mixture model correlation
%fits for active (red) and inactive stars (black). Stars with ambiguous categorisation are plotted as open circles. The measurements from our study
%(triangles) are labelled. For reference, we include the approximate activity level for WASP-51 expected from the stellar age (diamond); see
%section~\ref{Subsection:WASP-51} for details.}
%\label{Fig:Surfacegravity}
%\end{figure}

\begin{figure}
\includegraphics[width=85mm]{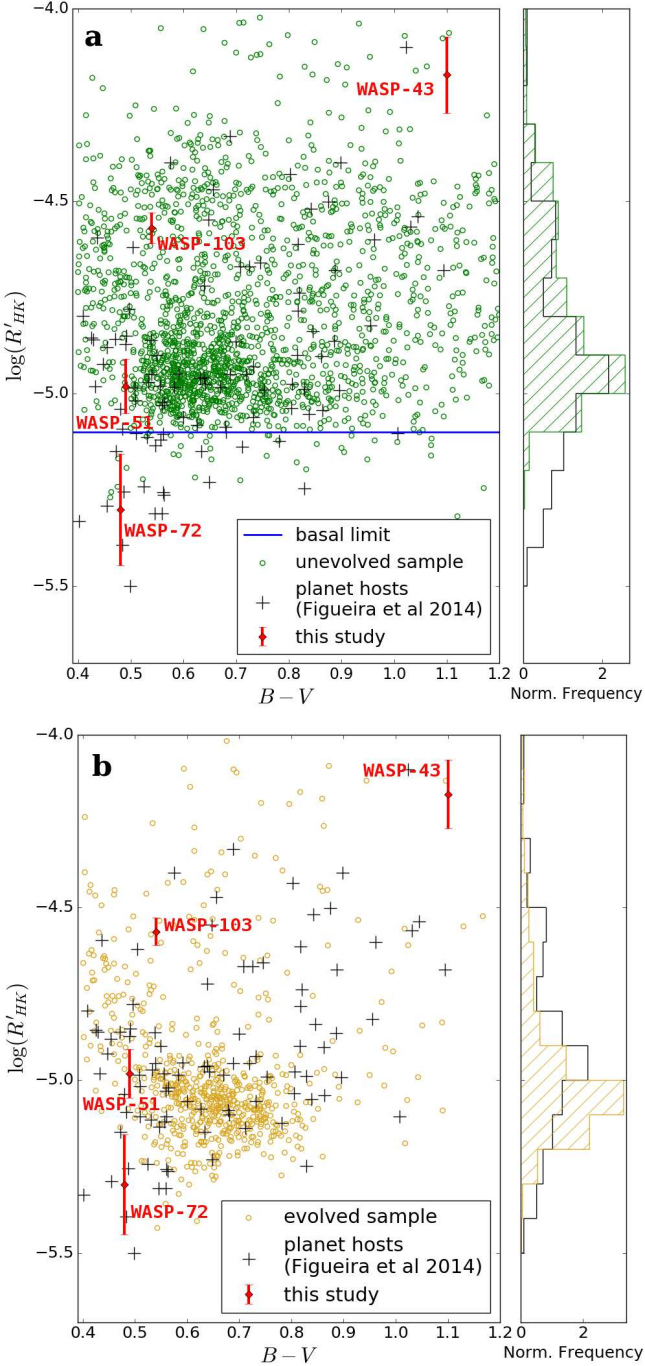}
%\caption{$\log({\rm \textit{R}'_{HK}})$ distribution for ,
%compared to archival data on planet hosts and our measurements.}

\caption{$\log({\rm \textit{R}'_{HK}})$ distribution for field stars compared to archival data on planet hosts and our measurements, using (a) unevolved, main sequence stars less than 0.45 mag above the average
main sequence
and (b) evolved field stars between 0.45 mag and 2 mag above the average main sequence. Note that bimodality in the distributions
is expected (e.g \citealt{Wright2004}, \citealt{Gray2006a}).
}
\label{Fig:Test1}
\end{figure}

Figure~\ref{Fig:Test1} shows our
new $\log({\rm \textit{R}'_{HK}})$ measurements in the context of
other transiting, short-period planet hosts
 \citep{Figueira2014}, and a large sample of field stars \citep{Pace2013}.
We limit all samples to $0.4 < $ B-V $< 1.2$ because $\log({\rm \textit{R}'_{HK}})$ is
well calibrated in this range and
%in the range of 0.4 - 1.0 \citep{Isaacson2010}, and
notably unreliable for B-V $>$ 1.2 \citep{Noyes1984}.
 %The latter category
%does not include known short period planets, although their presence cannot be excluded in most cases.
%For field stars in Figure~\ref{Fig:Test1}, we used the largest compilation of activity data published to date
Main sequence, subgiant and giant stars have very different activity distributions
(e.g. \citealt{Wright2004}; \citealt{Mittag2013}) so
it is useful to place stars on the HR diagram and differentiate them
by their evolutionary status.
We used absolute magnitudes and reliable B-V values from
%\citet{Pace2013} and
%entries with
the XHIP catalogue \citep{Anderson2012},
combined with the \citet{Wright2004} empirical average main sequence (MS), to calculate
the height above the MS for each star. $\log({\rm \textit{R}'_{HK}})$ was calculated from the XHIP B-V data and the mean \citet{Pace2013} \textit{S}-values.
We define an unevolved main sequence population
 %(up to a metallicity of [Fe/H] $\sim$ 0.3)
of stars less than 0.45 mag above the MS, as in \citet{Wright2004}.
A main sequence basal activity limit of $\log({\rm \textit{R}'_{HK}})$>-5.1 applies only to this sample, seen in Fig.~\ref{Fig:Test1}a.
Note this limit arises from chromospheric emission that is always present. It is the stellar analogue of the quiet sun chromospheric emission. The basal level is reached when stars are devoid of active regions (see e.g. \citealt{Schroder2012} and references therein).
%to obtain absolute magnitudes and reliable B-V values.

%Highly evolved stars follow a completely different colour-activity distribution than dwarfs and subgiants.
%Since
None of the planet hosts are highly evolved, so we reject
objects more than 2.0 mag above the MS, and define
a sample of moderately evolved stars between 0.45 and 2.0 mag above the MS, comprised
mostly of subgiants (Fig.~\ref{Fig:Test1}b).
%These have lower activity values than their unevolved counterparts.
For most planet hosts, precise parallaxes are unavailable, so we cannot distinguish between MS and moderately evolved stars in the same way.
We therefore compare planet hosts to both populations in Figure~\ref{Fig:Test1}.

\begin{figure}
\includegraphics[width=85mm]{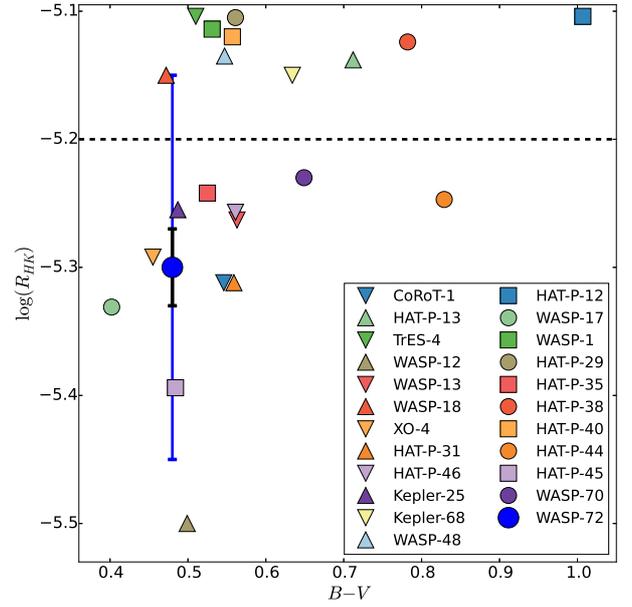}
\caption{Short period planet hosts from \citet{Figueira2014} with $\log({\rm \textit{R}'_{HK}})$ below the main sequence basal limit (-5.1) and
our measurement of WASP-72. For reference, we show the uncertainty with (blue) and without (black) the uncertainty in our calibration to 
the Mount Wilson system (Equation~2). Stars with $| \gamma - \rm{V}_{ISM} | < 15 \ \rm{km \ s}^{-1}$ (triangles) may 
be depressed below $\log({\rm \textit{R}'_{HK}})= -5.2$ by ISM absorption.}
\label{Fig:BelowBasal}
\end{figure}

These comparisons indicate that
%$\log({\rm \textit{R}'_{HK}})$ values for
transiting planet hosts have a greater spread in $\log({\rm \textit{R}'_{HK}})$ than field stars,
a statement which remains true whether
we consider main sequence or evolved stars.
We performed Anderson-Darling tests, comparing the $\log({\rm \textit{R}'_{HK}})$ distribution of the planet host
sample with our main sequence, evolved, and combined field star samples respectively. The resulting Anderson-Darling statistic
values of 27.7, 8.2 and 8.6 correspond to very low probabilities (3~x~$10^{-5}$, 4~x~$10^{-4}$, 3~x~$10^{-4}$) that the samples are drawn from the same parent
distribution in each case.
We find that the \citet{Figueira2014} dataset contains 22 planet hosts below the basal chromospheric emission limit (24 \% of the sample). Activity depression
%-by whatever mechanism- appears
%to be
is widespread in the known population of transiting
%short-period
planet hosts.
Only 2\% of our unevolved sample
%derived from \citet{Pace2013},
and 9\% of our combined field star sample, show
such an unusual activity level.
Fig.~\ref{Fig:BelowBasal} identifies
%We highlight
all planet hosts below the basal limit,
more than doubling the number previously identified \citep{Fossati2013}.
%Fig.~\ref{Fig:Test1}(a) shows that as stars evolve,
The horizontal basal envelope
which is fairly clear in the main sequence field star sample in Fig.~\ref{Fig:Test1}(a) becomes a diagonal envelope in Fig.~\ref{Fig:Test1}(b), with
higher $\log({\rm \textit{R}'_{HK}}) \sim -5.0$ for the bluest evolved stars and lower $\log({\rm \textit{R}'_{HK}}) \sim -5.2$
for the redder evolved stars (c.f. \citealt{Mittag2013}). WASP-72 and the outliers identified by \citet{Fossati2013} appear even more anomalous
in the context of Fig.~\ref{Fig:Test1}(b) than they do in the context of Fig.~\ref{Fig:Test1}(a).

\begin{figure}
\includegraphics[width=85mm]{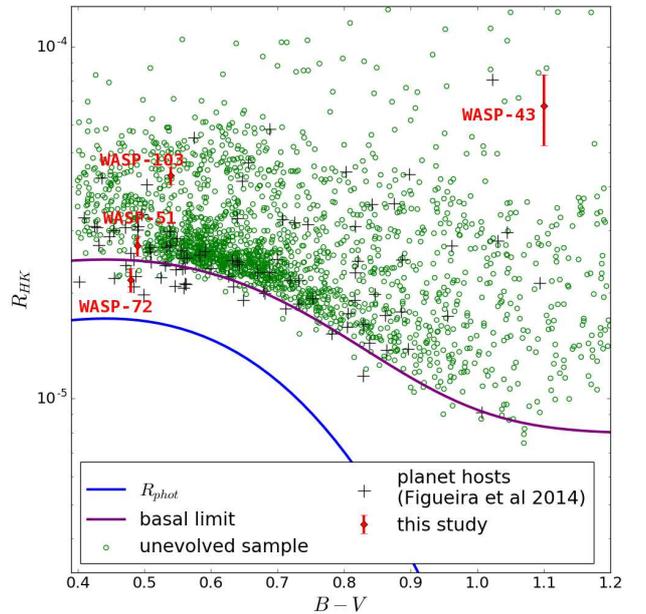}
%\caption{$\log({\rm \textit{R}'_{HK}})$ distribution for ,
%compared to archival data on planet hosts and our measurements.}

\caption{${\rm \textit{R}_{HK}}$ distribution for unevolved field stars and planet hosts,
compared to the chromospheric basal limit and ${\rm \textit{R}_{phot}}$, the photospheric contribution to ${\rm \textit{R}_{HK}}$.
}
\label{Fig:WithPhotospheric}
\end{figure}

Anomalously low $\log({\rm \textit{R}'_{HK}})$ is  seen for Kepler-25 and Kepler-68: two multi-planet systems each hosting 2 low-mass
(0.02 -0.08 $M_{\rm{J}}$), short period companions.
Tidal influences of these planets on their parent stars is negligible. Therefore, a tidal stellar activity suppression
mechanism as proposed for WASP-18 by \citet{Miller2012} can certainly be excluded in these cases.
Note that all anomalous planet hosts discussed fall below the chromospheric basal limit, but above ${\rm \textit{R}_{phot}}$, the purely photospheric
contribution to the \hbox{Ca {\sc ii} H\,\&\,K} bandpasses. This is shown in Figure~\ref{Fig:WithPhotospheric},
where we plot ${\rm \textit{R}_{phot}}$ as defined by \citet{Noyes1984} and ${\rm \textit{R}_{HK}} ={\rm \textit{R}_{phot}} + {\rm \textit{R'}_{HK}}$.

Transiting planet hosts are generally more distant than field stars with $\log({\rm R'_{HK}})$ measurements. Thus an alternative explanation for
$\log({\rm \textit{R}'_{HK}})$ depression in planet hosts is high absorption in the ISM along the line of sight.
%to a planetary system.
In the case of WASP-12, this is unlikely \citep{Fossati2013}, but \citet{Fossati2015}
found that the ISM explanation is viable for WASP-13, while circumstellar absorption could also be present.
To assess potential ISM absorption contributions for the planet hosts of interest, we calculated
the difference between the systems' radial velocities, $\gamma$, and the velocities of known local ISM clouds, $\rm{V}_{ISM}$ (Table~\ref{Table:ISMvelocity}).
We used the \citet{Redfield2008} ISM model\footnote{http://lism.wesleyan.edu/LISMdynamics.html},
identifying ISM clouds traversed along the relevant lines of sight and calculating the clouds' projected radial velocities.
ISM absorption can depress $\log({\rm \textit{R}'_{HK}})$ values if the velocity differences are within the \hbox{Ca {\sc ii} H\,\&\,K} core bandpass
widths ($\pm$ 86 km/s). The effect is much more pronounced for smaller $| \gamma - \rm{V}_{ISM} |$, amplified by
the triangular $H$ and $K$ bandpass weighting (Fossati et al, in prep.).
If we assume typical ISM column densities of $\rm{log} \ N_{CaII} = 12$ \citep{Welsh2010} for the relevant stellar distances (100-500 pc)
and conservatively assume a star has intrinsic chromospheric emission exactly at the basal level,
our calculations show that $| \gamma - \rm{V}_{ISM} | \lesssim 15 \ \rm{km \ s}^{-1}$ is required
to cause $\log({\rm \textit{R}'_{HK}})$ values lower than -5.2 (Fossati et al, in prep.).
We indicate $\log({\rm \textit{R}'_{HK}})=-5.2$ in Fig.~\ref{Fig:BelowBasal}, and differentiate objects with $| \gamma - \rm{V}_{ISM} | \lesssim 15 \ \rm{km \ s}^{-1}$ 
from the remainder. Objects with $\log({\rm \textit{R}'_{HK}}) < -5.2$ and $| \gamma - \rm{V}_{ISM} | > 15 \ \rm{km \ s}^{-1}$ are those most 
likely to be exhibiting significant circumstellar absorption of the \hbox{Ca {\sc ii} H\,\&\,K} line cores.
Using the assumptions outlined above and the $| \gamma - \rm{V}_{ISM} |$ values in Table~\ref{Table:ISMvelocity},
we expect values of the interstellar contribution to the $\log({\rm \textit{R}'_{HK}})$ depression for individual stars to be $\sim$ 0.03 - 0.12 dex.

Table~\ref{Table:ISMvelocity} and the line in Fig.~\ref{Fig:BelowBasal} is only indicative of potential ISM contributions
from known local clouds, mapped at low spatial resolution.
 A detailed case by case analysis (see \citealt{Fossati2013} and \citealt{Fossati2015}) is
needed for all anomalous planet hosts to assess the ISM versus circumstellar absorption contributions. This needs to take
into account the likely intrinsic activity levels in all cases, and
is beyond the scope of the current work.

\begin{table*}
 \centering
 \begin{minipage}{140mm}
  \caption{Differences between the stellar radial velocities ($\gamma$) and projected velocities of known ISM clouds \citep{Redfield2008}, traversed along the line-of-sight to each planet host
 below the chromospheric basal limit, and WASP-51. $\gamma$ values were taken from the Exoplanet Orbit Database and the individual planet discovery papers.}
  \begin{tabular}{@{}llll@{}}
  \hline
   System & $\gamma$ ($\rm{km \ s}^{-1}$) &  $| \gamma - \rm{V}_{ISM} |$ ($\rm{km \ s}^{-1}$) & ISM clouds  \\
  \hline
CoRoT-1&23.8 &3.2, 0.6 & LIC, Aur  \\
HAT-P-12&-40.6 & 34.7 & NGP \\
HAT-P-13&14.8& 0.3 & LIC \\
HAT-P-29&-21.7& 36.8 & LIC \\
HAT-P-31&-2.4& 14.7, 21.2, 26.5 & LIC, Mic, Oph\\
HAT-P-35&41.0& 24.2, 20.9 & LIC, Aur\\
HAT-P-38&-19.7& 37.4, 33.2 & LIC, Hya \\
HAT-P-40&-25.0 &  & none \\
HAT-P-44&-33.5& 27.5 & NGP \\
HAT-P-45&23.9& 55.3, 50.6, 63.5 & Oph, G, Aql\\
HAT-P-46&-20.9& 10.0, 6.5, 21.5 & Oph, G, Aql\\
Kepler-25&-8.5& 2.9, 10.5, 5.0 & LIC, Mic, G\\
Kepler-68&-20.9& 13.2, 5.8 & LIC, Mic \\
TrES-4&-15.9& 1.8, 4.3 & LIC, Mic \\
WASP-1&-13.4& 22.6 & LIC \\
WASP-12&18.9& 3.2 & LIC\\
WASP-13&9.9& 2.7 & LIC \\
WASP-17&-49.3& 21.0 & G \\
WASP-18&2.8& 2.0, 27.3, 8.4 & LIC, Dor, Cet \\
WASP-48&-19.7& 14.2 & LIC\\
WASP-70&-65.4& 43.6 & Mic \\
WASP-72&35.9& 23.4, 17.8 & LIC, G \\
XO-4&1.6 & 14.2 & LIC\\
\hline
WASP-51&44.7& 28.0, 24.8 & LIC, Aur \\
\hline
\end{tabular}
\label{Table:ISMvelocity}
\end{minipage}
\end{table*}

\subsection{Findings for Individual Systems}\label{Section:IndividualSystems}
%In the following, we discuss our activity measurements in the context of our 4 systems' parameters,
%focussing particularly on the stellar ages in the literature.

%\begin{table*}
% \centering
% \begin{minipage}{80mm}
%  \caption{Comparison of stellar ages in the literature with activity ages derived from the MH08 relation, using our $\log({\rm \textit{R}'_{HK}})$ values.}
%  \begin{tabular}{@{}lllll@{}}
%  \hline
%   System &   $\log({\rm \textit{R}'_{HK}})$ & activity age\footnote{Uncertainty propagated from uncertainty in  $\log({\rm \textit{R}'_{HK}})$. Note MH08 reports activity-age scatter at the 60\% and 30 \% levels for ages below and above 130 Myr respectively.}  & literature age & reference   \\
%  \hline
%  WASP-43          & -4.17 $\pm$ 0.10 & 45 $\pm$ 40 Myr & 300-600 Myr & \citealt{Hellier2011} \\
%  WASP-51 & -4.98 $\pm$ 0.07 & 6.2 $\pm$ 1.3 Gyr & 0.5-1.8 Gyr  & \citealt{Johnson2011} \\
%                                                     & & & 0.6-1.0 Gyr & \citealt{Bonfanti2015}  \\
%  WASP-103         & -4.57 $\pm$ 0.04 \footnote{mean of our measurements} & 950 $\pm$ 230 Myr & 1.8-6.2 Gyr & \citealt{Southworth2014} \\
%                                                     & & & 3.0-5.0 Gyr & \citealt{Gillon2014}  \\
%  WASP-72          & -5.30 $\pm$ 0.15 & 11 $\pm$ 0.8 Gyr & 2.6-3.8 Gyr  & \citealt{Gillon2013} \\

%\hline
%\end{tabular}
%\label{Table:AgeComparison}
%\end{minipage}
%\end{table*}

\subsubsection{WASP-43}

WASP-43b orbits
%is one of the few HJs discovered around a relatively late,
a K-type star with $T_{\rm{eff}}$ = 4520 K \citep{Hellier2011} and is consequently less irradiated than most HJ planets with similar orbital proximity (c.f. Table~\ref{Table:BasicData}).
WASP-43 has a rotation period of 15.6 d, i.e. a gyrochronological age of 300-600 Myr
\citep{Hellier2011}. This is consistent with its X-ray emission \citep{Czesla2013},
but tidal spin-up could affect both, obscuring the true system age.
Our $\log({\rm \textit{R}'_{HK}})$ value places WASP-43 at the activity level seen for very young ($<$~130 Myr; MH08) members of the
Pleiades cluster and Sco-Cen association in Figure~\ref{Fig:Test1}.
WASP-43 lies well above the Hyades,  $\sim$ 625 Myr,  activity level
at this spectral type; see Fig.4 in \citet{Paulson2002}.
The $\log({\rm \textit{R}'_{HK}})$-age relation of MH08 gives a very low age (Table \ref{Table:HostActivities}), inconsistent with WASP-43's gyrochronological age.
The activity value is not significantly affected by erroneous extrapolation of our calibration to large S-values.
To verify this, we compared our spectrum of WASP-43 with
%From the \citet{Pace2013} catalogue, we selected a comparison star with basic stellar parameters close to WASP-43's values.
%The star with the highest activity value that meets these criteria is HD\,86856 ,
%measured from
a HIRES spectrum of a similar star, HD\,86856, with $\log({\rm \textit{R}'_{HK}})$ = -4.37 and
%   We degraded the archival spectrum to our RSS resolution; the comparison
confirmed that WASP-43's core emission is higher by the amount expected from our $\log({\rm \textit{R}'_{HK}})$ value.

MH08 found a tight correlation between stellar X-ray emission, parameterised by
$\log({\rm \textit{R}_{X}})$, and $\log({\rm \textit{R}'_{HK}})$.
Using WASP-43's $\log({\rm \textit{R}_{X}}) = -4.98 \pm 0.23$ \citep{Czesla2013}, equation A1 in MH08 predicts $\log({\rm \textit{R}'_{HK}}) = -4.56 \pm 0.07$.
Our measured value, $\log({\rm \textit{R}'_{HK}})= -4.17 \pm 0.10 $ constitutes a clear outlier with respect to Figure 15 of MH08, far beyond the scatter among
$\sim$ 200 stars therein.
The X-ray activity level agrees with the stellar rotation period, and  our $\log({\rm \textit{R}'_{HK}})$ measurement is
anomalously high. This could be due to a stellar flare during the exposure producing  an increase in
\hbox{Ca {\sc ii} H\,\&\,K} emission by a
factor of 2 to 3.
Such short-term variability
has been observed for the young, rapidly rotating K5V star BD+201790, which hosts a disputed HJ
(\citealt{Hernan-Obispo2010}; \citealt{Hernan-Obispo2015}).
However, stars similar to WASP-43 with $ 4000\,{\rm K} < T_{\rm{eff}} <  5000\,{\rm K  \, ; \,}$ and $10 {\rm d} < P_{ \rm{rot}} < 20 {\rm d}$ have
flare frequencies of
0.04 ${\rm d}^{-1}$ - 0.21 ${\rm d}^{-1}$ with durations of  0.1 - 15 hr  \citep{Balona2015}.
This implies a $< 14\%$ chance of seeing a flare during our 0.56 hr exposure even at the most favourable flare frequencies and durations.

%In this context, our
%measurement is in dramatic disagreement with \citep{Czesla2013}.
%Using equation A1 from MH08, we would expect , given the $\log({\rm \textit{R}_{X}})$ value.
%This discrepancy at the 0.4 dex level constitutes a clear outlier in Given that clearly anomalously high.
%It is possible that this is due to a stellar flare
%during the exposure. If the X-ray derived $\log({\rm \textit{R}'_{HK}})$ level represents a quiescient baseline, this implies  given our uncertainties in $\log({\rm \textit{R}'_{HK}})$.
%Spectroscopic flare observations for late K-stars, covering the \hbox{Ca {\sc ii} H\,\&\,K} lines are extremely rare in the literature.
%Short-timescale variability in excess of this level

Our measurement was taken at WASP-43\,b
orbital phase 0.66, whereas the X-ray data from \citet{Czesla2013} covers the secondary eclipse of the planet.
\citet{Shkolnik2008} observed clear \hbox{Ca {\sc ii} H\,\&\,K} emission increases for the
HJ host HD\,179949 at phases 0.6 - 1 and attributed this to SPI. If WASP-43 shows
similar behaviour, our data could be affected by SPI, whereas the X-ray data is not.
The putative strong flaring event during our exposure could have been induced by SPI increasing the
flaring frequency above that of stars without HJ companions.
Only mixed evidence exists for \hbox{Ca {\sc ii} H\,\&\,K} emission increases due to SPI at the few \% level
\citep{Miller2015}.
The doubling or tripling implied by our measurement is unique, and requires confirmation.

\begin{figure}
\includegraphics[width=84mm]{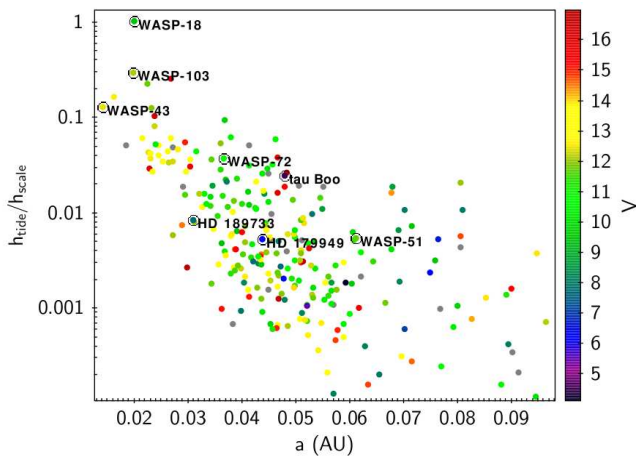}
\caption{SPI interaction strength proxies for planets with $M_{\rm{p}}$sin\textit{i} = 0.1-13 $M_{\rm{J}}$ and $a <$ 0.1 AU. The visual magnitude is shown
as a guide to the observability of the systems. Planets observed in this study and key systems from the literature are labelled.
A null-result for SPI in the particularly extreme WASP-18 system was found by \citet{Miller2012}, and attributed to the very
low intrinsic activity level of this F6V star; see X-ray nondetection in \citet{Pillitteri2014}. Evidence for weak SPI effects have been reported for
HD\,179949 (\citealt{Shkolnik2003}), HD\,189733 \citep{Poppenhaeger2014a} and tau Boo \citep{Walker2008}.
}
\label{Fig:SPIproxies}
\end{figure}

Figure~\ref{Fig:SPIproxies} shows all planets from the Exoplanet Orbit Database with $0.1 M_{\rm{J}} < M_{\rm{p}}$, $\sin\textit{i} < 13 M_{\rm{J}}$
and $a <$ 0.1 AU,
the hot Jupiters observed in this paper and several systems where evidence for SPI has been published.
We plot the tidal SPI proxy $h_{tide}/h_{scale}$, following \citet{Cuntz2000},
where $h_{tide}$ is the height of the tidal bulge raised on the star,
and $h_{scale}$ is photospheric scale height (see also e.g. \citealt{Poppenhaeger2014a}; \citealt{Pillitteri2014}).
In the absence of detailed theoretical modelling and magnetic field maps, simple scaling proxies based on known system properties are the only available guide to assess
putative magnetic SPI strength.
Ideally, the maximum power released by magnetic reconnection would be calculated analytically following \citet{Lanza2013}.
 A meaningful estimate requires measurements of the stellar field strength and configuration, which are not available
for our targets.
It is worth noting that \citet{Miller2015} use $M_{\rm{p}}$sin\textit{i}$/ a^{2}$ as a magnetic SPI strength proxy,
highlighting potential activity enhancement in ``extreme'' systems with $M_{\rm{p}}$ sin\textit{i}$/ a^{2}$ > 450 (see their Fig.~7.)
WASP-103 and WASP-43 have $M_{\rm{p}}$ sin\textit{i}$/ a^{2}$ values 8 and 20 times larger than this threshold.
However, this is not an ideal proxy to differentiate magnetic from tidal SPI effects: $M_{\rm{p}}$sin\textit{i} strongly correlates with $h_{tide}/h_{scale}$
due to the $M_{\rm{p}}$sin\textit{i} dependence. It is also questionable to assume all close-in planets of interest orbit in the regime
where the stellar field strength falls of as $a^{-2}$, since this will change with the geometry of the fields (see e.g.
\citealt{Lanza2013}). We choose to simply plot the semi-major axis in Figure~\ref{Fig:SPIproxies} as an indicator of
magnetic interaction strength. There is insufficient information about the stellar field environments for the vast majority
of exoplanets to differentiate them into subsets of other parameterisations.
Both proxies plotted are exceptionally favorable for the WASP-43 system. The alternative simple proxies used in the literature, such as $M_{\rm{p}}$sin\textit{i}$/ a^{2}$,
$M_{\rm{p}}$sin\textit{i}$/ a$ and $M_{\rm{p}}$sin\textit{i}$/P$ (\citealt{Miller2015}; \citealt{Poppenhaeger2010}; \citealt{Shkolnik2008})
also all indicate SPI is particularly likely for WASP-43.  Note that WASP-43b and WASP-103 are the Hot Jupiters
with the shortest and 4th shortest semi-major axis discovered to date.

Since our activity value is anomalously high with respect to the stellar rotation
period, a tidal explanation (invoking enhanced rotationally-driven emission) is disfavored for WASP-43.

%$\log({\rm \textit{R}'_{HK}})$  for WASP-43 is anomalously high.
We find no evidence for circumstellar absorption in accordance with \citet{Salz2015}'s finding that  WASP-43\,b should be stable against mass loss.
% may be negligible.
Time-variable magnetic SPI at an unprecedented level could be an explanation for our high value of $\log({\rm \textit{R}'_{HK}})$.
To distinguish this explanation from ``normal'', flare activity, phase-resolved
monitoring of X-ray and/or optical activity indicators of WASP-43 is needed.

\subsubsection{WASP-51/HAT-P-30}\label{Subsection:WASP-51}
This planet
%was discovered independently in 2011 by the HATNET and SWASP surveys (\citealt{Johnson2011};
%\citealt{Enoch2011}) and is therefore referred to as either HAT-P-30 or %WASP-51.
%It
is a typical spin-orbit misaligned HJ (\citealt{Johnson2011}; \citealt{Enoch2011}).
% hot Jupiter
%and has received limited dedicated follow-up to date.
% give the
The most precise literature ages (Table \ref{Table:HostActivities}) are based on
isochrone dating,
using the stellar parameters
of \citet{Johnson2011}.
The high stellar Li abundance is consistent with a $\leq$ 1 Gyr age \citep{Enoch2011}.
Our $\log({\rm \textit{R}'_{HK}})$ derived age (Table \ref{Table:HostActivities}) is clearly anomalously high.
%Taken as a whole,
In summary, it seems the true age is $\sim$ 1 Gyr and the normal stellar activity
has been depressed,
either through the planet tidally suppressing the stellar magnetic dynamo
%normal stellar activity
%(c.f. \citealt{Miller2012} and \citealt{Pillitteri2014})
or through absorption of the \hbox{Ca {\sc ii} H\,\&\,K} emission by circumstellar gas from the planet or by the ISM.
%(c.f. \citealt{Haswell2012}; \citealt{Fossati2013}).
The MH08 relation predicts
%WASP-51 should exhibit
$\log({\rm \textit{R}'_{HK}}) \sim$ -4.6 at a 1 Gyr age.
%We have plotted this value in Fig.~\ref{Fig:Surfacegravity} to highlight the discrepancy.
%This high intrinsic activity results in the observed, depressed value remaining above the basal limit.

If the intrinsic chromospheric emission is indeed at that level, our measurement implies
depression of $\log({\rm \textit{R}'_{HK}})$ by $\sim$ 0.4 dex. This is unlikely to arise
from absorption in the ISM: our calculations show that clouds with $| \gamma - \rm{V}_{ISM} | \lesssim 30 \ \rm{km \ s}^{-1}$ and
a column density $\rm{log} \ N_{CaII} > 16$ would be required (Fossati et al, in prep). For stars out to 800 pc
\citet{Welsh2010} consistently observe $\rm{log} \ N_{CaII} < 13$ (typically $\sim$ 12), and
WASP-51 lies at $\sim$190 pc \citep{Johnson2011}.

Note that the mean activity level reported in \citet{Johnson2011} using the HIRES spectrograph is $S_{\rm{MW}}$ = 0.128 $\pm$ 0.014
%or$\log({\rm \textit{R}'_{HK}})$ = -5.24 $\pm$ 0.19.
%This is the mean value from
%their observations over a timespan of 180 days, showing little variability.
The uncertainty is taken to be 11 \%, as reported for the HIRES calibration to the Mount Wilson system \citep{Isaacson2010}.
The difference between this $S_{\rm{MW}}$ value and our measurement (Table~\ref{Table:HostActivities}) is
at the 1.5 $\sigma$ level, i.e. not significant.

\subsubsection{WASP-103}
WASP-103b is an extreme HJ,
%;in terms of irradiation level and expected tidal effects
%\citep{Gillon2014}. Orbital decay is predicted to be measurable over the next decade \citep{Birkby2014},
%and it is one of the most tidally distorted planets known
close to tidal disruption (\citealt{Gillon2014}; \citealt{Southworth2014}).
%Among our 4 targets, it is the most similar system to the extreme case of WASP-12b.
%The recent discovery of WASP-103b means there have been few follow-up studies to date.
%
The $S_{\rm{RSS}}$ RMS variability over our 16 observations of WASP-103 is at the 2 \% level,
comparable to
%not significantly larger than
the photon noise uncertainties.
We saw
%divided all spectra obtained by the spectrum with highest SNR, and confirmed there were
no significant changes
in the \hbox{Ca {\sc ii} H\,\&\,K} line core profiles, though
the data sample a narrow phase range away from transit, so we are unable to address the issue of phase-dependent absorption.
%as seen in WASP-12 \citep{Fossati2010; Haswell2012}. %do not sample a wide phase range, and no observations in or near transit
%were obtained.
The dispersion in $\log({\rm \textit{R}'_{HK}})$ is 0.02. This is lower than the uncertainties quoted
for individual values, since those include additional systematic calibration and B-V uncertainties.

Our mean $\log({\rm \textit{R}'_{HK}})$ value of -4.57 corresponds to a low activity-age compared with alternative estimates
(Table \ref{Table:HostActivities}), but the
significance of the
age difference is only high with respect to the \citet{Gillon2014} age estimate (3 $\sigma$). Improved stellar parameters are needed for a better constrained
independent age assessment (c.f. Table 6 in \citealt{Southworth2014}).
By analogy with WASP-12 (\citealt{Fossati2010}; \citealt{Haswell2012}) we expected depressed activity due to absorption by circumstellar gas from the extreme HJ planet.
Either this is masked by SPI or the planetary mass loss rate may be unexpectedly low. Modelling of atmospheric escape, as for WASP-43, would be useful to investigate this possibility.
Figure~\ref{Fig:SPIproxies} suggests SPI enhancements are plausible.
WASP-103 has the second-highest tidal interaction proxy value $h_{tide}/h_{scale}$ of all known exoplanets. The value is a third of the extreme WASP-18 system, for which
the possibility of tidal activity suppression rather than enhancement has been invoked \citep{Pillitteri2014}.
%Clearly, there is no evidence for a depressed activity value compared to both the main sequence basal level
%and the literature ages of WASP-103 shown in Table \ref{Table:HostActivities}. This is a surprising result, given the similarities to the WASP-12 system.
%The activity value is somewhat higher than expected for the system age, but the
%WASP-103 may be significantly affected by SPI, as its interaction strength proxy values fall amongst the highest of all hot Jupiters
%as shown in Figure~\ref{Fig:SPIproxies}. This could mask activity depression from absorption. Alternatively

\subsubsection{WASP-72}
WASP-72 is among the most highly irradiated hot Jupiters known \citep{Gillon2013}.
%, experiencing approximately 60\% of the WASP-103b irradiation level.
%There have been no follow-up studies of this planet to date.
%This is the only target where
Our measured activity value falls below the basal level, see Figure~\ref{Fig:Test1} and \ref{Fig:BelowBasal}.
%joining the
%anomalous cluster of hot Jupiter hosts in .
While the uncertainty on $\log({\rm \textit{R}'_{HK}})$ is relatively large,
WASP-72 certainly falls amongst the lowest activity
outliers for both main sequence and evolved stars of its spectral type.
In Figure~\ref{Fig:BelowBasal} we show the substantial uncertainty contribution from 
the calibration to the Mount Wilson system. Note that intercomparing activity values
measured with the same instrument setup does not suffer this systematic uncertainty.
$\log({\rm \textit{R}'_{HK}})$ implies WASP-72 is far older than stellar evolution models suggest
(Table~\ref{Table:HostActivities}, MH08), strengthening the case for a depressed activity value. We conclude WASP-72 is a system
where relatively weak \hbox{Ca {\sc ii} H\,\&\,K} emission
from an inactive, $\sim$3 Gyr old star \citep{Gillon2013} is absorbed sufficiently to appear below the basal limit,
analogously to other close-in planet systems (\citealt{Lanza2014}; \citealt{Fossati2015a}).
With currently available data for the WASP-72 system, we cannot disentangle circumstellar absorption
from planetary mass loss and absorption in the ISM. However, our estimates in Section~\ref{Section:ActivityContext}
indicate the known ISM clouds along the line of sight (Table~\ref{Table:ISMvelocity}) cannot entirely explain the anomalously low value even if the
intrinsic stellar emission is exactly at the basal level.

\section{Conclusions}\label{Section:Conclusion}

We calibrated the RSS at SALT to measure chromospheric activity on the Mount Wilson system.
We used this to measure the activity of four HJ host stars.
\citet{Fossati2013} highlighted the anomalously low $\log({\rm \textit{R}'_{HK}})$ values for WASP-12 and five other planet hosts included in
\citet{Knutson2010}. We
revisit this with the significantly extended dataset of \citet{Figueira2014}, finding that 24 \% of the sample (22 hosts) show these anomalies including two low mass,
multi-planet systems and WASP-72 (Figs.~\ref{Fig:Test1} and \ref{Fig:BelowBasal}).

%
%showed that the expanded dataset of \citealt{Figueira2014} greatly increases this anomalous planet host sample.
%It is no longer limited to hot Jupiter systems and contains two hosts of .
% WASP-72 joins these outliers in Fig.~\ref{Fig:Test1}, while
WASP-43 is an outlier in the opposite corner of Fig.~\ref{Fig:Test1},
%Half of our sample are clear outliers when compared with the field star populations.
while WASP-51/HAT-P-30 has an anomalously
low $\log({\rm \textit{R}'_{HK}})$ for its age.
%as suggested by other indicators.
There may be (at least) two
processes operating in close-in HJ systems which affect the observed $\log({\rm \textit{R}'_{HK}})$ values.
Anomalously high
$\log({\rm \textit{R}'_{HK}})$ values may be attributed to SPI \citep{Cuntz2000}.
%From consideration of the
%physical mechanisms likely to be important in these systems, we can attribute candidate mechanisms to both sets of outliers.
The outliers in the bottom left of Fig.~\ref{Fig:Test1} can be attributed to diffuse circumstellar gas lost from the planets which
absorbs the stellar chromospheric emission in strong resonance lines (\citealt{Haswell2012}; \citealt{Fossati2013}). Absorption by the ISM
may also play a role for some of these systems.
%WASP-72 and WASP-51 are both located away from the Galactic plane, at moderate distances (340 pc and 190 pc respectively; \citealt{Gillon2013},
%\citealt{Johnson2011}), so low interstellar column densities are expected towards them.
Our estimates suggest the latter cannot explain the anomalous values of WASP-51 and WASP-72.
Further investigations are needed to examine any interstellar and circumstellar
absorption contributions,
%as was done for WASP-12 \citep{Fossati2013}, WASP-18 \citep{Fossati2013} and WASP-13 \citep{Fossati2015},
and to test whether the intrinsic activity levels of WASP-72 and WASP-51 are indeed normal.

Further measurements are needed to investigate whether
WASP-43's anomalously high $\log({\rm \textit{R}'_{HK}})$ is time-dependent.
The extreme HJ WASP-103\,b also displays higher activity than expected from the system age;
improved
stellar parameters are needed to test this conclusion.
%In Fig.~\ref{Fig:SPIproxies} we show the distribution of two predictors of the strength of SPI.
WASP-43 and WASP-103  have exceptionally high predicted SPI levels from the simple scaling laws illustrated in Fig.~\ref{Fig:SPIproxies}.

A quarter of the host stars of short period exoplanets exhibit anomalously low values of $\log({\rm \textit{R}'_{HK}})$.
%This is the case both for our RSS measurements and the large sample reported by \citealt{Figueira2014}, where we found that a quarter of short period planet hosts
%fall below the basal limit.
%Low
Activity can appear depressed
%anomalies can arise from
by circumstellar absorption of the
stellar \hbox{Ca {\sc ii} H\,\&\,K} flux.
% by circumstellar gas. High anomalies may be due to
SPI can potentially enhance the normal \hbox{Ca {\sc ii} H\,\&\,K} emission.
In some systems, including WASP-103, both mechanisms may operate simultaneously.
Such systems would not necessarily have obviously anomalous $\log({\rm \textit{R}'_{HK}})$ activity ages.
We recommend that the $\log({\rm \textit{R}'_{HK}})$ index is never used as an age indicator for stars which host
close-in planets: it is unreliable for this purpose.
Disentangling the mechanisms to the extent of being able to predict which one dominates will require extensive
uniform observational work, covering a range of stellar spectral types, ages, planet masses,
planet radii, Roche lobe radii, orbital separations, and measurements of the stellar magnetic field configurations.

%With a sample of the latter systems, we could test our speculative suggestion that WASP-103's activity is
%boosted by SPI.
%Clearly, the use of $\log({\rm \textit{R}'_{HK}})$ activity ages
%for such stars should be avoided.

\section*{Acknowledgments}

All of the observations reported in this paper were obtained with the Southern African Large Telescope (SALT), in which the Open University is a shareholder,
as part of the UK SALT Consortium.
We thank the SALT astronomers for making our observations, Geoff Bradshaw for IT support, in particular for OU undergraduate GS, and CH thanks M. Williams and R. Gandhi-Patel for logistical support.
DS is supported by an STFC studentship, CAH and JRB are supported by STFC under grant ST/L000776/1. 
This research has made use of the SIMBAD database,
operated at CDS, Strasbourg, France.

\bibliographystyle{mnras}
\bibliography{RSSpaperReferences-final}

\bsp

\label{lastpage}

\end{document}